# ukbFGSEA: an R package for applying fast preranked gene set enrichment analysis to UK Biobank exome data

**Pengjun Guo[1,2] and He Zhu[3]**

1 SDSZ International Department, Beijing, China

2 The Experimental High School attached to Beijing Normal University, Beijing, China

3 University College London, London, UK

## Abstract

Motivation

The Genebass dataset, released by Karczewski et al. (2022), provides a comprehensive resource elucidating associations between genes and 4,529 phenotypes based on nearly 400,000 exomes from the UK Biobank. This extensive dataset enables the evaluation of gene set enrichment across a wide range of phenotypes, facilitating the inference of associations between specified gene sets and phenotypic traits. Despite its potential, there is currently no established method for applying gene set enrichment analysis (GSEA) to Genebass data.

Results

To address this gap, we propose utilizing fast preranked gene set enrichment analysis (FGSEA) as a novel approach to determine whether a specified set of genes is significantly enriched in phenotypes within the UK Biobank. We developed an R package, ukbFGSEA, to implement this analysis, complete with a hands-on tutorial. Our approach has been validated by analyzing gene sets associated with autism spectrum disorder, developmental disorder, and neurodevelopmental disorders, demonstrating its capability to reveal both established and novel associations.

Availability and implementation

The ukbFGSEA source code and tutorial are freely available at https://github.com/AzuleneG/ukbFGSEA

Contact: pengjunguo@outlook.com, hedyzhu615@gmail.com

**Introduction**

The release of the Genebass dataset[1] marks a significant milestone in genetic research. By elucidating associations between genes and 4,529 phenotypes from nearly 400,000 exomes in the UK Biobank, Genebass offers a comprehensive resource of unprecedented scale and quality. This dataset provides a unique opportunity to explore the genetic underpinnings of numerous diseases and traits, making it an invaluable tool for researchers aiming to uncover causal relationships and novel biological insights.

One powerful approach for extracting meaningful information from such a vast dataset is gene set enrichment analysis (GSEA)[2]. GSEA is a robust method for investigating the correlation between predefined sets of genes and various phenotypes. By examining the collective behavior of groups of genes, GSEA can uncover significant biological contexts and reveal intricate gene-phenotype correlations. However, despite the potential of Genebass, there is currently no established method for applying GSEA to this rich dataset.

Addressing this gap is crucial for advancing our understanding of complex diseases and their genetic basis. Therefore, we propose utilizing fast pre-ranked gene set enrichment analysis (FGSEA)[3] as a novel approach to identify significant associations between specified gene lists and phenotypes within the UK Biobank. To facilitate this analysis, we have developed an R package, ukbFGSEA, specifically designed to integrate seamlessly with the Genebass dataset.

Our package, ukbFGSEA, is equipped with a hands-on tutorial, enabling users to apply FGSEA to Genebass data efficiently. It supports two methods for ranking gene sets and generates easily interpretable enrichment plots and computational results. By leveraging the extensive resource provided by Genebass, ukbFGSEA empowers researchers to perform detailed evaluations of gene set enrichment across a wide range of phenotypes. This capability enhances our ability to infer associations between specific gene sets and phenotypic traits, ultimately advancing our understanding of the genetic architecture of complex diseases.

## Implementation

### Genebass Dataset

The original Genebass gene-level burden results in HT format were downloaded from Genebass. We extracted gene-level results for all 4,529 phenotypes from the HT file and stored them in RDS format for accessibility and efficiency in R. The RDS file can be downloaded from Google Drive https://drive.google.com/file/d/1j0bAXm2AYmZNQxQBbFODien0dpKUGCHn/view?usp=drive_link.

### Functions in the R Package

The ukbFGSEA package contains three main functions:

1. ukbfgsea

This primary function applies FGSEA to gene-level burden results from a single phenotype with a single variant functional annotation. The function begins by preparing the Genebass data, which involves filtering out any NA values to ensure data integrity. Next, two ranking metrics are calculated to prioritize genes based on their statistical significance and effect sizes:

- Beta Ranking: The beta ranking metric $f_{beta}$ is computed as $f_{beta} = -log_{10}(p) \times \beta$. This metric combines the significance of the association (p-value) and the direction and magnitude of the effect size (beta), providing a comprehensive measure for ranking genes.

- Sign Ranking: The sign ranking metric $f_{sign}$ is computed as $f_{sign} = -log_{10}(p) \times sign(\beta)$. This metric emphasizes the direction of the effect, irrespective of its magnitude, which can be useful in certain biological contexts where the direction of the association is more critical than its size.

The ranked data sets are then processed using the `fgseaMultilevel()` function from the `fgsea` package, chosen for its efficiency and accuracy in handling large-scale datasets like those from Genebass.

2. map_ukbfgsea

This function extends the application of `ukbfgsea` to multiple phenotypes, enabling users to run FGSEA on hundreds of thousands of phenotypes simultaneously.

3. tidy_map

This function processes the raw output from `map_ukbfgsea()`, organizing the results into a tidy format. This involves splitting the output into distinct columns for each phenotype and each ranking method, providing a clear and accessible summary of the FGSEA results.

Comprehensive documentation, including a hands-on tutorial, is available on the package website https://github.com/AzuleneG/ukbFGSEA. This documentation ensures users can effectively apply the package to their research.

**Use case**

To demonstrate the utility of `ukbFGSEA`, we applied the `map_ukbfgsea()` function to three gene sets associated with autism spectrum disorder (ASD185), developmental disorder (DD477), and neurodevelopmental disorders (NDD664) as described by Fu et al. (2022)[4]. These gene sets provide a robust basis for validating the `ukbFGSEA` approach and exploring novel associations.

Methods

The case study began by filtering out phenotypes categorized as 'icd_first_occurrence,' which represent the date of the first occurrence of a trait or phenotype. We then excluded phenotypes with fewer than 3,000 cases in the UK Biobank to ensure sufficient statistical power. The analysis focused on loss-of-function (LoF) mutations to target the most impactful genetic variants. This process yielded a well-powered set of 1,603 phenotypes for further analysis.

Using `map_ukbfgsea()`, FGSEA was conducted across the filtered Genebass phenotypes for each gene set. The results were then cleaned and organized using `tidy_map()`. This process is depicted in Figure 1, which illustrates the application of our methodology to the ASD185 gene set as example.

**Results**

We retained phenotypes with significant p-values from both ranking metrics for each gene set,

1. ASD185: Among the 36 phenotypes enriched with ASD genes, well-documented associations such as epilepsy and panic attacks reinforce known neurological links[5,6]. Notably, novel associations with medical procedures including subacromial decompression, carpal tunnel, and knee replacement surgeries indicate emerging areas of concern within physical health that are possibly linked to ASD. Moreover, the inclusion of severe medical conditions like myocardial infarction expands the recognized spectrum of comorbid health risks[7]. These novel findings underscore the need for further investigation to grasp their full implications on ASD management, potentially inspiring more holistic healthcare strategies.
2. DD477: The 116 phenotypes enriched with DD genes span from established to novel associations. Traditional links with conditions like epilepsy and cataracts echo known neurological and sensory impacts[8-10]. However, new associations with complex medical treatments such as subacromial decompression and interventions for inflammatory bowel disease suggest unexplored genetic pathways that may influence both developmental and systemic health.
3. NDD664: Analysis of the 164 phenotypes enriched with NDD genes showcases their extensive influence, reaffirming connections with conditions like epilepsy and stroke[10-14] and introducing novel associations with procedures such as subacromial decompression. These insights suggest uncharted areas where NDD genes may affect physical health, highlighting the necessity for ongoing research into these connections. The identification of links to arthritis and complex therapies also opens new avenues for understanding the comprehensive impact of NDD genes, potentially catalyzing advancements in both diagnostics and therapeutic interventions.

Collectively, these findings illuminate the complex interplay between these genes and a broad spectrum of health outcomes. By confirming established comorbidities and unveiling unexpected correlations, our analysis not only deepens the understanding of these disorders' genetic bases but also reveals the extensive roles these genes may play across various physiological and medical domains. Future research is imperative to decipher the specific mechanisms underlying these associations. Such insights are crucial for developing more precise diagnostic tools and targeted interventions, ultimately enhancing healthcare delivery for individuals grappling with these conditions.

**Conclusion**

The ukbFGSEA R package effectively applies FGSEA to the Genebass dataset, bridging a critical methodological gap. By validating known associations and discovering new ones, ukbFGSEA demonstrates its potential to advance our understanding of complex phenotypes in the UK Biobank. The tool is robust, user-friendly, and capable of generating significant genetic insights.

**Acknowledgments**

We thank Jinjie Duan for testing the software and insightful discussions.

**Conflict of interest**

None declared.

**Data availability**

All data and code utilized in the analyses presented in this article are available on GiHub at https://github.com/AzuleneG/ukbFGSEA. Additionally, the genebass dataset, including gene-level burden test results for all phenotypes, is accessible via Google Drive at

https://drive.google.com/file/d/1j0bAXm2AYmZNQxQBbFODien0dpKUGCHn/view?usp=drive_link.

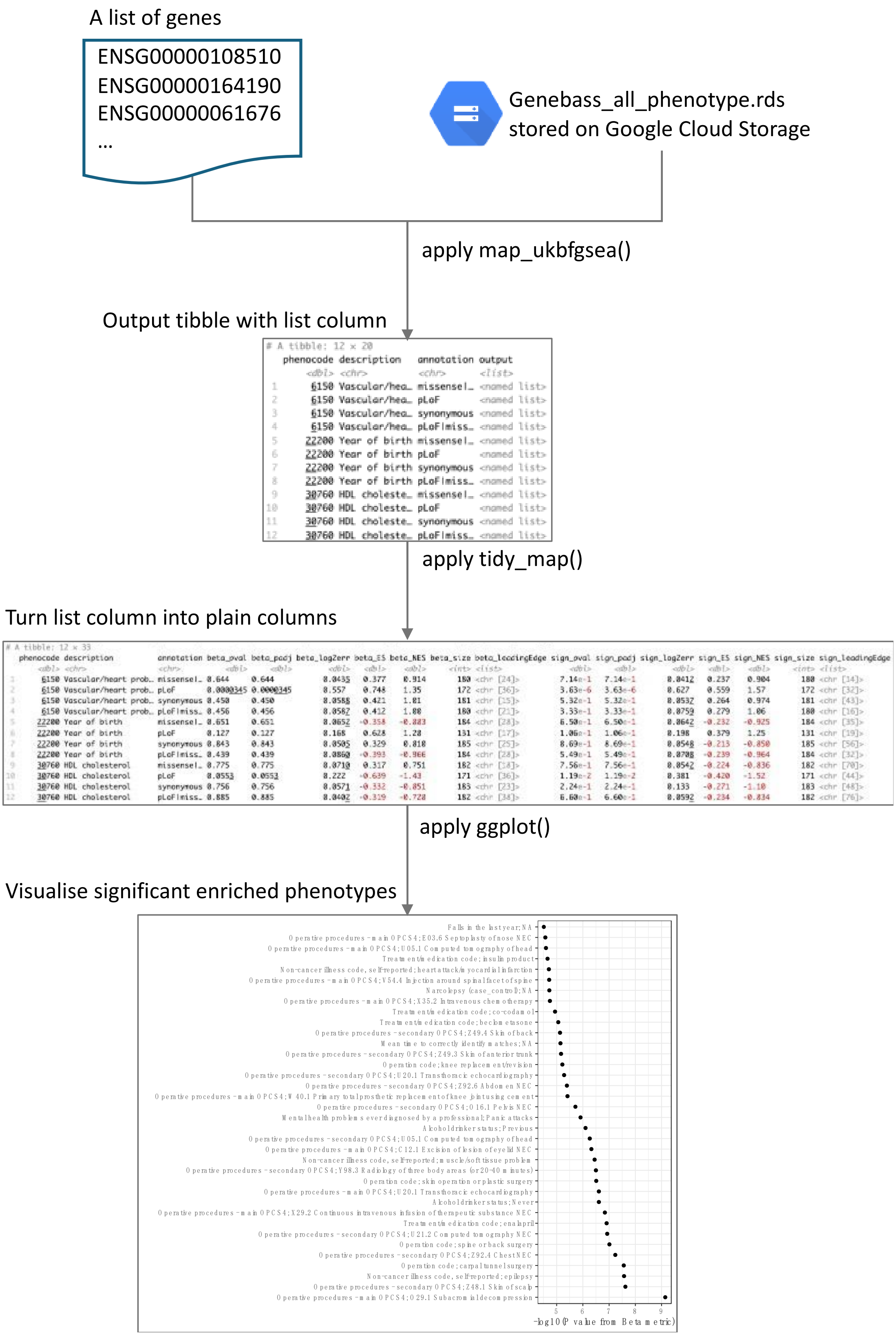

Figure 1. The application of the ukbFGSEA R package to the ASD185 gene set within the Genebass dataset. It illustrates the steps from gene list input to phenotypic enrichment analysis, showcasing

how users can leverage the ukbFGSEA package to explore genetic enrichment within the UK Biobank. The bottom plot depicts -log10 p-values from the Beta metric, indicating the strength of each gene-phenotype association. The R code utilized for conducting this analysis is freely available at the tutorial https://github.com/AzuleneG/ukbFGSEA